\documentclass[showpacs,preprint,amsmath,amssymb,nofootinbib]{revtex4-1}

\usepackage{amsmath}
\usepackage{graphicx}
\usepackage{epsf}
\usepackage{psfrag}
\usepackage{epsfig}
\usepackage{graphics}
\usepackage{amsfonts}
\usepackage{epstopdf}
\usepackage{slashed}

\begin{document}
\newcommand{\e}{\mbox{e}}
\newcommand{\be}{\begin{equation}}
\newcommand{\ee}{\end{equation}}
\newcommand{\bq}{\begin{eqnarray}}
\newcommand{\eq}{\end{eqnarray}}
\newcommand{\intx}{\int^{1}_{0} dx}
\newcommand{\pbruto}{\hbox{$p \!\!\!{\slash}$}}
\newcommand{\eps}{\epsilon}
\newcommand{\qbruto}{\hbox{$q \!\!\!{\slash}$}}
\newcommand{\lbruto}{\hbox{$l \!\!\!{\slash}$}}
\newcommand{\kbruto}{\hbox{$k \!\!\!{\slash}$}}
\newcommand{\bc}{\begin{center}}
\newcommand{\ec}{\end{center}}
\newcommand{\dd}{\frac{d^2k}{(2 \pi)^2}}
\newcommand{\dt}{\frac{d^3k}{(2 \pi)^3}}
\newcommand{\dtp}{\frac{d^3p}{(2 \pi)^3}}
\newcommand{\dq}{\frac{d^4k}{(2 \pi)^4}}
\newcommand{\dn}{\frac{d^nk}{(2 \pi)^n}}
\newcommand{\PLB}{{\it{Phys. Lett. {\bf{B}}}}}
\newcommand{\NPB}{{\it{Nucl. Phys. {\bf{B}}}}}
\newcommand{\PRD}{{\it{Phys. Rev. {\bf{D}}}}}
\newcommand{\AOP}{{\it{Ann. Phys. }}}
\newcommand{\MPL}{{\it{Mod. Phys. Lett. }}}

\newcommand{\del}{\partial}
\newcommand{\delbruto}{\hbox{$\partial \!\!\!{\slash}$}}
\newcommand{\elemint}{\int \frac{d^4 k}{(2\pi)^4} \ }
\newcommand{\intk}{\int_{k} }
\newcommand{\kumbruto}{\hbox{$k \!\!\!{\slash}_{1}$}}
\newcommand{\kdoisbruto}{\hbox{$k \!\!\!{\slash}_{2}$}}
\newcommand{\kum}{\hbox{$k_{1}$}}
\newcommand{\kdois}{\hbox{$k_{2}$}}
\newcommand{\Tr}{\mathrm{Tr}}
\newcommand{\kummu}{\hbox{$k_{1\mu}$}}
\newcommand{\kdoismu}{\hbox{$k_{2\mu}$}}
\newcommand{\kumnu}{\hbox{$k_{1\nu}$}}
\newcommand{\kdoisnu}{\hbox{$k_{2\nu}$}}
\newcommand{\Ilog}{I_{log}^{(1)}(\lambda^2) }
\newcommand{\Ilogd}{I_{log}^{(2)}(\lambda^2) }
\newcommand{\Iquad}{I_{quad}^{(1)}(\lambda^2) }
\newcommand{\Iquadd}{I_{quad}^{(2)}(\lambda^2) }
\newcommand{\lnk}{\mathrm{ln} \bigg(-\frac{k^{2}}{\lambda^{2}} \bigg)}
\newcommand{\lnkum}{\mathrm{ln} \bigg(-\frac{(k+k_{1})^{2}}{\lambda^{2}} \bigg)}
\newcommand{\lnkdois}{\mathrm{ln} \bigg(-\frac{(k+k_{2})^{2}}{\lambda^{2}} \bigg)}
\newcommand{\lnkumdois}{\mathrm{ln} \bigg(-\frac{(k_{1}-k_{2})^{2}}{\lambda^{2}} \bigg)}

\title{Momentum Routing Invariance in Feynman Diagrams and Quantum Symmetry Breakings}

\date{\today}

\author{L. C. Ferreira$^{(a)}$} \email[]{luellerson@fisica.ufmg.br}
\author{A. L. Cherchiglia$^{(a)}$} \email[]{adriano@fisica.ufmg.br}
\author{Brigitte Hiller$^{(b)}$} \email[]{brigitte@teor.fis.uc.pt}
\author{Marcos Sampaio$^{(a)}$} \email []{msampaio@fisica.ufmg.br}
\author{M. C. Nemes$^{(a)}$}\email[]{mcnemes@fisica.ufmg.br}

\affiliation{(a) Universidade Federal de Minas Gerais - Physics
Department - ICEx \\ P.O. BOX 702, 30.161-970, Belo Horizonte MG -
Brazil}
\affiliation{(b) Physics Department -
Centre for Science and Technology\\
Coimbra University\\
Rua Larga, P-3004 516 Coimbra - Portugal}
\begin{abstract}

\noindent

We illustrate with examples that quantum symmetry breakings in perturbation theory are connected to breakdown of momentum routing invariance (MRI) in the loops of a Feynman diagram. We show that MRI is a necessary and sufficient condition to preserve abelian gauge symmetry at arbitrary loop order. We adopt the implicit regularization framework in which surface terms that are directly connected to momentum routing can be constructed to arbitrary loop order. The interplay between momentum routing invariance, surface terms and anomalies is discussed. We also illustrate that MRI is important to preserve supersymmetry. For theories with poor symmetry content, such as scalar field theories, MRI is shown to be important in the calculation of renormalization group functions. 
\end{abstract}


\maketitle

\section{Introduction}

In the 1972 seminal paper by 't Hooft and Veltman \cite{thooftveltman} on dimensional regularization (DReg), they emphasized that besides respecting unitarity and causality, DReg also allowed for shifts in integration variables (loop momenta) of Feynman amplitudes. When no quantum symmetry breakings occurred, Ward identities were automatically satisfied crowning DReg as the ideal framework to handle ultraviolet divergences in perturbation theory of gauge theories. Indeed it is well known that the possibility of shifts in the integration variable is an important ingredient for diagrammatic proofs of gauge invariance in quantum electrodynamics.

In the early eighties, motivated by the construction of a framework applicable to models which are incompatible with dimensional continuation on the space-time dimension (for instance supersymmetric, chiral and topological quantum field theories), Elias, McKeon, Mann and collaborators \cite{MCKEON} brought back the problem on loop momentum routing ambiguities. The latter stem from shift of integration variable surface terms which appear in the integer dimension but not in dimensionally continued space-times.
Such ambiguities were used by the authors to warrant the validity of Ward-Slavnov-Taylor  identities in some model calculations at one-loop level. In other words, this approach, called Pre-regularization, used integration variable ambiguities in four dimensional loop integrals to parametrize the divergent amplitudes in a way that Ward identities were preserved by  a suitable choice of the routing labels in the loop of a diagram. Anomalies, such as the well known Adler-Bardeen-Bell-Jackiw triangle chiral anomaly, appear in this approach when the ambiguities proved themselves insufficient to preserve the full set of symmetry identities valid at classical level.

Of course Dimensional Reduction (DRed) is the most popular tool to perform
Feynman diagram calculations in supersymmetric gauge theories and
other dimension specific models which require modifications in
gauge symmetry preserving dimensional regularization. Generalizations of
DRed which ensure invariance to two-loop order in models of phenomenological importance have been done
\cite{Stockinger}, but it is still unknown to what extent it preserves
supersymmetry. An {\emph{invariant}}
regularization framework , which avoids the task of computing
symmetry restoring counterterms order by order in perturbation theory for dimensional sensitive theories is hitherto unknown.
In other words, the construction of an invariant regularization scheme to compute with supersymmetric gauge theories is
justified on practical and theoretical grounds.

The main purpose of this contribution is to argue that any regularization framework that operates on the physical dimension of the theory is invariant (that is to say, does not lead to quantum symmetry breakings of classical symmetries) if momentum routing invariance (MRI) in the Feynman diagrams is respected. The reverse of the coin is that anomalies manifest themselves as breaking of momentum routing invariance, which has been conjectured before by Jackiw in \cite{Jackiw:1999qq}-\cite{JACKIWCURRENTALGEBRA}. Moreover we argue that Implicit Regularization (IReg) is the ideal arena to formulate and illustrate this idea to arbitrary loop order, and therefore can be an useful tool for Feynman diagram calculations when dimensional methods become dodgy.

This work is organized as follows. In section \ref{secir} we present an overview of Implicit Regularization. The relationship
between momentum routing invariance and surface terms is shown in section \ref{secmr}. We present a link between MRI and abelian gauge invariance
to all loop orders in section \ref{secqed}. We also discuss the interplay between momentum routing breaking and anomalies in the context of the Adler-Bardeen-Bell-Jackiw triangle chiral anomaly. In section \ref{secex} we study
the role played by MRI in two contexts. We verify that supersymmetry is preserved in the Wess-Zumino model in component fields
only if MRI is respected. In addition, for quantum field theories with low symmetry content such as $\phi_{6}^{3}$ theory, we verify
that the universality of the first two renormalization scheme independent terms of the $\beta$-function holds, provided MRI is respected.
 We leave
section \ref{secco} to our concluding remarks.

\section{Aper\c{c}u on Implicit Regularization}
\label{secir}

Implicit Regularization \cite{Mota} is a momentum space framework which assumes tacitly that  a general
(multi-loop) Feynman amplitude is regularized only to allow  for a
mathematical identity at the level of propagators be used to
display its divergent content as basic divergent integrals (BDI's)
in terms of one internal momentum only. For a $n$-loop amplitude one can reduce $n-1$ internal momenta by judiciously subtracting subdivergencies.

The BDI's can be
absorbed in the definition of renormalization constants without being explicitly evaluated.
This strategy generalizes straightforwardly to higher loop order
because IReg is compatible with the local version of the BPHZ forest
formula based on the subtraction of local counterterms
\cite{Bogoliubov} \cite{Carlos2}, \cite{Edson}, \cite{Cleber} and
hence complying with locality, Lorentz invariance and unitarity \cite{Adriano}.
The same set of BDI's can be used to describe
the ultraviolet divergent content of both  massive and massless
quantum field theoretical models. A generalization of this procedure to handle
infrared divergences was constructed \cite{BJP}, \cite{BETAYM}.
An arbitrary positive (renormalization group) mass scale $\lambda$,
 appears via a regularization independent identity
which enables us to write a BDI as a function of $\lambda$ only
plus logarithmic functions of $\mu/\lambda$, $\mu$ being a
fictitious mass which is added to massless propagators. The limit
$\mu \rightarrow 0$ is well defined for the whole amplitude if it
is power counting infrared convergent ab initio.

It is important to notice that neither BDI's nor their derivatives with
respect to $\lambda$, which are also expressible as BDI's and are
useful to compute renormalization group functions, need be
evaluated. For instance, in the case when the ultraviolet behaviour is logarithmical, a
$n$-loop Feynman amplitude is cast as a finite
function of external momenta plus a set of BDI's, say
$I_{log}^{(i)} (\lambda^2)$, $i=1,\ldots,n$ plus {\emph{surface terms}}
expressed by integrals in $d$-dimensional space-time of a total
derivative in momentum space. The origin of these surface terms are
differences between $I_{log}^{(i)} (\lambda^2)$ and $I_{log}^{(i)
\mu_1 \mu_2 ...} (\lambda^2)$ where the latter is a
logarithmically divergent integral which contains in the integrand
a product of internal momenta carrying Lorentz indices $\mu_1,
\mu_2 ...$. In other words, in the process of isolating the
divergent content of an amplitude to loop integrals,
$I_{log}^{(i) \mu_1 \mu_2...} (\lambda^2)$ may be written as a
product of metric tensors symmetrized in the Lorentz indices times
$I_{log}^{(i)} (\lambda^2)$ (BDI) plus a surface term.  Such
local, {\emph{ regularization dependent}} surface terms are intrinsically
arbitrarily valued. It was shown, initially to one-loop order, that setting them to zero
corresponds to invoking translational invariance of Green's
functions \cite{Mota}. A constrained version of implicit regularization (CIReg) in which surface terms are systematically
set to zero was conjectured to preserve gauge symmetry and supersymmetry \cite{Edson}, \cite{Prd1}, \cite{Prd2}, \cite{David},
\cite{Eloy2}, although the link with MRI was not made clear.

It is interesting to compare these features with gauge invariant
DReg and Constrained Differential Renormalization (CDfR) \cite{DiffRen2}.
On one hand, IReg surface terms explicitly evaluated using DReg
vanish which in turn also allows for shifts in the integration
variable at zero cost. On the other hand the rules of CDfR which
deliver gauge invariant amplitudes at one-loop level translate into momentum space
as freedom of shifts in integration variable and discard of
surface terms. In \cite{Carlos1} we have shown the correspondence
between CIReg and CDfR in one-loop level. CDfR was constructed  for
one-loop calculations although differential renormalization can be applied to multi-loop
graphs in principle. Generalization of CDfR to two-loop order is
discussed in \cite{Seijas}. Because the surface terms in IReg are
well known, the generalization to multi-loop amplitudes in gauge
theories is clear \cite{Edson}.

The idea of BDI's which
underlies the ultraviolet divergent content of a general $n$-loop
Feynman diagram can be extended to represent infrared divergences as well.
This is important because in dimensional methods
ultraviolet and infrared divergences are mixed. The latter appear
in perturbation theory in many fashions. Unlike ultraviolet divergences which are removed by
renormalization, infrared divergences are expected to cancel if a
correct definition of physical processes is made, but an intermediate regularization is needed. Moreover in supergraph
approach to supersymmetric models, along with on-shell infrared
divergences of Yang-Mills theory, additional infrared divergences
appear \cite{Abbott} which must be {\emph{consistently}} separated
from ultraviolet ones before renormalization is effected.

Here we present a few formulae which will be useful in the next sections. We restrict ourselves to massless theories because the BDI's for  massive theories are exactly the same. In other words, defining the renormalization constants in terms of basic divergent integrals only corresponds to a mass independent renormalization scheme.
Suppose we have a n-loop Feynman amplitude with $L$ external legs. In order to express its divergent content in terms of a BDI in one-loop momentum only we need to perform $n-1$ integrations. The order in which they are performed can be chosen systematically to clearly display the counterterms to be subtracted in compliance with the Bogoliubov's recursion formula \cite{Adriano}. Having made this choice, the general form of the terms of the Feynman amplitude after $l$ integrations is
\begin{align}
&I^{\nu_{1}\ldots \nu_{m}}\!=\!\!\int\limits_{k_{l}}\!\frac{A^{\nu_{1}\ldots \nu_{m}}(k_{l},q_{i})}{\prod_{i}[(k_{l}-q_{i})^{2}-\mu^{2}]}\ln^{l-1}\!\left(\!-\frac{k_{l}^{2}-\mu^{2}}{\lambda^{2}}\right)\!,
\label{I}
\end{align}
\noindent
where $l=1, \cdots , n$. In the following, we are assuming an even number of dimensions. We can deduce a similar expression for odd dimensions as well.

In the above equation, $\int_{k_{l}}\equiv\int d^dk_{l}/(2 \pi)^d$, $q_{i}$ is an element (or combination of elements) of the set $\{p_{1},\ldots,p_{L},k_{l+1},\ldots,k_{n}\}$, and $\mu^{2}$ is an infrared regulator. Since the original integral is infrared safe, the limit $\mu^2\rightarrow 0$ is well defined and must be taken in the end of the calculation. The logarithmical dependence appears because this is the characteristic behaviour of the finite part of massless amplitudes \cite{Delamotte}. $\lambda^{2}$ is an arbitrary non-vanishing parameter with dimension of mass which parametrizes the freedom one has to subtract the divergences (renormalization group scale). It appears at one-loop level and survives to higher orders through a regularization independent mathematical identity (eq. \ref{scale}) as we show in the end of this section. The function $A^{\nu_{1}\ldots \nu_{m}}(k_{l},q_{i})$ is proportional to all possible combinations of $k_{l}$ and $q_{i}$ compatible with the Lorentz structure.

Assuming that a regulator $\Gamma$ is implicit in (\ref{I}), we may use the following identity in the denominators
\begin{align}
\frac{1}{(k_{l}-q_{i})^2-\mu^2}=\sum_{j=0}^{n_{i}^{(k_{l})}-1}\frac{(-1)^{j}(q_{i}^2-2q_{i} \cdot k_{l})^{j}}{(k_{l}^2-\mu^2)^{j+1}}
+\frac{(-1)^{n_{i}^{(k_{l})}}(q_{i}^2-2q_{i} \cdot k_{l})^{n_{i}^{(k_{l})}}}{(k_{l}^2-\mu^2)^{n_{i}^{(k_{l})}}
\left[(k_{l}-q_{i})^2-\mu^2\right]},
\label{ident}
\end{align}
in which the values of $n_{i}^{(k_{l})}$ are chosen such that all divergent integrals have a denominator free of $q_{i}$. Therefore, we can write the divergent integrals as a combination of
\begin{align}
I_{log}^{(l)}(\mu^2)&\equiv \int\limits_{k_{l}}^{\Lambda} \frac{1}{(k_{l}^2-\mu^2)^{d/2}}
\ln^{l-1}{\left(-\frac{k_{l}^2-\mu^2}{\lambda^2}\right)},\quad
\label{Ilogilog}\\
I_{log}^{(l)\nu_{1} \cdots \nu_{r}}(\mu^2)&\equiv \int\limits_{k_{l}}^{\Lambda} \frac{k_{l}^{\nu_1}\cdots
k_{l}^{\nu_{r}}}{(k_{l}^2-\mu^2)^\beta}
\ln^{l-1}{\left(-\frac{k_{l}^2-\mu^2}{\lambda^2}\right)},
\label{IlogLorentz}
\end{align}
\noindent
or
\begin{align}
I_{quad}^{(l)}(\mu^2)&\equiv \int\limits_{k_{l}}^{\Lambda} \frac{1}{(k_{l}^2-\mu^2)^{\frac{d-2}{2}}}
\ln^{l-1}{\left(-\frac{k_{l}^2-\mu^2}{\lambda^2}\right)},\quad
\label{Iquadiquad}\\
I_{quad}^{(l)\nu_{1} \cdots \nu_{r+2}}(\mu^2)&\equiv \int\limits_{k_{l}}^{\Lambda} \frac{k_{l}^{\nu_1}\cdots
k_{l}^{\nu_{r+2}}}{(k_{l}^2-\mu^2)^\beta}
\ln^{l-1}{\left(-\frac{k_{l}^2-\mu^2}{\lambda^2}\right)},
\label{IquadLorentz}
\end{align}
\noindent
where $r=2\beta-d$. It is important to note that only these type of divergences appear because linear BDI's always vanish. Although we have already reduced the divergences to basic divergent integrals free of external momenta, we can show that the integrals defined above are related. For example, in a case with two Lorentz indices we have
\begin{align}
&I_{log}^{(l)\,\mu
\nu}(\mu^2)=\sum_{j=1}^{l}\left(\frac{2}{d}\right)^j\!\frac{(l-1)!}{(l-j)!}\!\left\{\frac{g^{\mu \nu}}{2}I_{log}^{(l-j+1)}(\mu^2)-\frac{1}{2}\Upsilon_{0}^{(l)\,\mu\nu}\right\},
\label{identsurface1}
\end{align}
\noindent
where $\Upsilon_{0}^{(l)\,\mu\nu}\equiv \int\limits_{k}\!\!\frac{\partial}{\partial k^{\mu}}\left[\frac{k^{\nu}}{(k^2-\mu^2)^{d/2}}\ln^{l-j}\!{\left(-\frac{k^2-\mu^2}{\lambda^2}\right)}\!\right]$, and
\begin{align}
&I_{quad}^{(l)\,\mu
\nu}(\mu^2)=\sum_{j=1}^{l}\left(\frac{2}{d-2}\right)^j\!\frac{(l-1)!}{(l-j)!}\!\left\{\frac{g^{\mu \nu}}{2}I_{quad}^{(l-j+1)}(\mu^2)-\frac{1}{2}\Upsilon_{2}^{(l)\,\mu\nu}\right\},
\label{identsurface}
\end{align}
\noindent
where $\Upsilon_{2}^{(l)\,\mu\nu} \equiv \int\limits_{k}\!\!\frac{\partial}{\partial k^{\mu}}\left[\frac{k^{\nu}}{(k^2-\mu^2)^{\frac{d-2}{2}}}\ln^{l-j}\!{\left(-\frac{k^2-\mu^2}{\lambda^2}\right)}\!\right]$.

\noindent
In the previous equations, $\Upsilon_{0}^{(l)\,\mu\nu}$, and $\Upsilon_{2}^{(l)\,\mu\nu}$ are (arbitrary) regularization dependent surface terms. In a more general case they are given by
\begin{align}
\Upsilon_{i}^{(l)\nu_{1}\cdots\nu_{j}}\equiv\int{\frac{d^{d}k}{(2 \pi)^d}}\frac{\partial}{\partial k_{\nu_{1}}}\frac{k^{\nu_{2}}\cdots k^{\nu_{j}}}{(k^{2}-\mu^{2})^{\frac{d+j-2-i}{2}}}\ln^{l-1}\Bigg[-\frac{(k^{2}-\mu^{2})}{\lambda^{2}}\Bigg].
\label{tsdef}
\end{align}
\noindent
An equivalent definition in terms of Lorentz scalar objects $\Gamma_{i}^{(l,j)}$ is
\begin{align}
g^{\{\nu_{1}\cdots\nu_{j}\}}\Gamma_{i}^{(l,j)}\equiv\Upsilon_{i}^{(l)\nu_{1}\cdots\nu_{j}},
\label{idets}
\end{align}
\noindent
where $g^{\{\nu_{1}\cdots\nu_{j}\}}\equiv g^{\nu_{1}\nu{2}}\cdots g^{\nu_{j-1}\nu_{j}}+$ symmetric combinations.

Finally the divergences can be written in terms of (\ref{Ilogilog}) and (\ref{Iquadiquad}). However from (\ref{Ilogilog}) we see that this integral is ultraviolet and infrared divergent as $\mu^2\rightarrow 0$. In order to separate these divergences and define a genuine ultraviolet divergent object we use the regularization independent relation
\begin{align}
I_{log}^{(l)}(\mu^2)&=I_{log}^{(l)}(\lambda^2)-\frac{b_{d}}{l}\ln^{l}\left(\frac{\mu^2}{\lambda^2}\right)+b_{d}\sum_{k=1}^{A}\binom{A}{k}\sum_{j=1}^{l-1}\frac{(-1)^k}{k^j}\frac{(l-1)!}{(l-j)!}\ln^{l-j}\left(\frac{\mu^2}{\lambda^2}\right),
\label{scale}\\
\mbox{where}\quad\lambda^{2}&\neq0,\;\; A\equiv\frac{(d-2)}{2} \mbox{,} \;\; b_{d}\equiv\frac{i}{(4\pi)^{d/2}}\frac{(-1)^{d/2}}{\Gamma(d/2)}.\label{bd}
\end{align}

For infrared safe models the infrared divergence must cancel in the amplitude as a whole. This in fact occurs because, as we use identity (\ref{ident}), the finite part of the amplitude will also have a logarithmical dependence in $\mu^2$ which exactly cancels the infrared divergence coming from the use of the scale relation  (\ref{scale}).

This procedure was shown to comply with unitarity, locality and Lorentz invariance in \cite{Adriano}. The whole program is compatible with overlapping divergences through the Bogoliubov's recursion formula which means that the divergent content of an arbitrary Feynman graph can always be cast as a basic divergent integral.

\section{Momentum Routing Invariance and Surface Terms at $n$-loop order}
\label{secmr}

In this section we study the conditions which guarantee MRI to an arbitrary multi-loop Feynman diagram. We find that the only condition needed to preserve such symmetry is to set surface terms to zero.

As it is well known, if $f(k)$ is an arbitrary function, then
\begin{align}
f(k+a)&=f(k)+a_{\sigma}\frac{\partial}{\partial k_{\sigma}}f(k)+\frac{a_{\sigma}a_{\rho}}{2!}\frac{\partial}{\partial k_{\sigma}k_{\rho}}f(k)+\cdots\nonumber\\
&=exp\left(a_{\sigma}\frac{\partial}{\partial k_{\sigma}}\right)f(k).
\label{expansion}
\end{align}
\noindent
Moreover if $B\equiv \int{\frac{d^{d}k}{(2 \pi)^d}}\frac{\partial f(k)}{\partial k^{\sigma}}$ and  $f(k)$ is finite or logarithmic divergent, then by Gauss' theorem $B$ is null.

\begin{figure}[ht!]
\begin{center}
\includegraphics{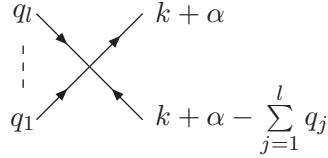}
\end{center}
\vspace{-0.5cm}
\caption{\label{figvertex} Generic vertex with arbitrary momentum routing $\alpha$.}
\end{figure}

We now consider an arbitrary graph at one-loop order. Setting $k$ as its internal momentum, and $q_{i}$ as the external momenta we will have in general, for any theory, a vertex of the type depicted in figure \ref{figvertex}.

\noindent
Therefore, the amplitude of this graph can be expressed as
\be
A\equiv\int{\frac{d^{d}k}{(2 \pi)^d}}f(k+\alpha,q_{i}),
\ee
\noindent
where for simplicity we consider a scalar amplitude, although the generalization for amplitudes with an arbitrary number of Lorentz indices is straightforward.
We now present the cornerstone of our argument: if we have momentum routing invariance then
\be
\int{\frac{d^{d}k}{(2 \pi)^d}}\left[f(k+\alpha,q_{i})-f(k+\beta,q_{i})\right]=0
\ee
\noindent
must be satisfied for arbitrary $\alpha$ and $\beta$. Using identity (\ref{expansion}) it reduces to
\be
\int{\frac{d^{d}k}{(2 \pi)^d}}\left[exp\left(\alpha_{\sigma}\frac{\partial}{\partial k_{\sigma}}\right)-exp\left(\beta_{\sigma}\frac{\partial}{\partial k_{\sigma}}\right)\right]f(k,q_{i})=0.
\label{condition}
\ee
\noindent
After space-time and internal group algebra are performed, $f(k,q_{i})$ is given by
\be
f(k,q_{i})=\frac{g(k,q_{i})}{\prod\limits_{j=1}^{L}\left[(k+l_{j}(q_{i}))^2-\mu^2\right]}
\ee
\noindent
where $g(k,q_{i})$ and $l_{j}(q_{i})$ are polynomials in the momenta and $\mu^2$ is an infrared regulator. The divergence of such integral is controlled by the dimension of the theory ($d$), the number of internal lines ($L$), and the degree in $k$ of the polynomial $g(k,q_{i})$ which we define as $m$.

Evidently, if $d+m-2L\leq0$ condition (\ref{condition}) is automatically satisfied as logarithmic divergent graphs are always momentum routing invariant. We proceed to linear divergent integrals which, after using the identity of IReg in all propagators one time, furnishes:
\be
f(k,q_{i})=f_{lin}(k,q_{i})+f_{log}(k,q_{i})+f_{fin}(k,q_{i}).
\ee
In view of the comments above only the first term is of interest to us. Its general form is given by
\be
\int{\frac{d^{d}k}{(2 \pi)^d}}f_{lin}(k,q_{i})=\int{\frac{d^{d}k}{(2 \pi)^d}}\frac{\prod_{i}(k\cdot q_{i})^{b_{i}}\prod_{k}(q_{i}\cdot q_{k})^{c_{ik}}}{\left[k^2-\mu^2\right]^{L}},
\ee
\noindent
where $d+s-2L=1$, $s\equiv\sum_{i}b_{i}$, and we cancelled powers of $k^2$ in the numerator against propagators. This cancellation must always be performed because, as we are dealing with divergent integrals, symmetric integration is forbidden \cite{Carlos1}, \cite{PerezVictoria:2001ej}.

In this case, condition (\ref{condition}) is satisfied only if
\be
(\alpha_{\sigma}-\beta_{\sigma})\int{\frac{d^{d}k}{(2 \pi)^d}}\frac{\partial}{\partial k_{\sigma}}f_{lin}(k,q_{i})=(\alpha_{\sigma}-\beta_{\sigma})h_{\nu_{1}\cdots\nu_{s}}(q_{i})\int{\frac{d^{d}k}{(2 \pi)^d}}\frac{\partial}{\partial k_{\sigma}}\frac{k^{\nu_{1}}\cdots k^{\nu_{s}}}{\left[k^2-\mu^2\right]^{L}}=0.
\ee
\noindent
Since $\alpha$, $\beta$ are arbitrary, and $h_{\nu_{1}\cdots\nu_{s}}(q_{i})$ is a polynomial in $q_{i}$, we notice that the condition above is equivalent to the statement (see equation (\ref{tsdef}))
\be
\Upsilon_{0}^{(1)\sigma\nu_{1}\cdots\nu_{s}}=0.
\ee

In other words, momentum routing invariance is guaranteed for linearly divergent graphs only if surface terms are systematically set to zero. We consider now quadratically divergent integrals which, after using identity (\ref{ident}) in all propagators two times, gives:
\be
f(k,q_{i})=f_{quad}(k,q_{i})+f_{lin}(k,q_{i})+f_{log}(k,q_{i})+f_{fin}(k,q_{i}).
\ee
Since linear, logarithmical and finite cases were already analysed, we only need to deal with the first term which is
\be
\int{\frac{d^{d}k}{(2 \pi)^d}}f_{quad}(k,q_{i})=\int{\frac{d^{d}k}{(2 \pi)^d}}\frac{\prod_{i}(k\cdot q_{i})^{b_{i}}\prod_{k}(q_{i}\cdot q_{k})^{c_{ik}}}{\left[k^2-\mu^2\right]^{L}},
\ee
\noindent
where $d+s-2L=2$, and $s\equiv\sum_{i}b_{i}$.

Now, condition (\ref{condition}) is satisfied only if
\be
(\alpha_{\sigma}-\beta_{\sigma})\int{\frac{d^{d}k}{(2 \pi)^d}}\frac{\partial}{\partial k_{\sigma}}f_{quad}(k,q_{i})=(\alpha_{\sigma}-\beta_{\sigma})h_{\nu_{1}\cdots\nu_{s}}(q_{i})\int{\frac{d^{d}k}{(2 \pi)^d}}\frac{\partial}{\partial k_{\sigma}}\frac{k^{\nu_{1}}\cdots k^{\nu_{s}}}{\left[k^2-\mu^2\right]^{L}}=0,
\ee
\noindent
and
\begin{align}
(\alpha_{\sigma}&-\beta_{\sigma})(\alpha_{\rho}-\beta_{\rho})\int{\frac{d^{d}k}{(2 \pi)^d}}\frac{\partial^{2}}{\partial k_{\sigma}\partial k_{\rho}}f_{quad}(k,q_{i})=(\alpha_{\sigma}-\beta_{\sigma})(\alpha_{\rho}-\beta_{\rho})h_{\nu_{1}\cdots\nu_{s}}(q_{i})\times\nonumber\\\times\Bigg[&g^{\nu_{1}\rho}\int{\frac{d^{d}k}{(2 \pi)^d}}\frac{\partial}{\partial k_{\sigma}}\frac{k^{\nu_{2}}\cdots k^{\nu_{s}}}{\left[k^2-\mu^2\right]^{L}}+\sum\limits_{j=2}^{s-1}g^{\nu_{j}\rho}\int{\frac{d^{d}k}{(2 \pi)^d}}\frac{\partial}{\partial k_{\sigma}}\frac{k^{\nu_{1}}\cdots k^{\nu_{j-1}}k^{\nu_{j+1}}\cdots k^{\nu_{s}}}{\left[k^2-\mu^2\right]^{L}}+\nonumber\\&+g^{\nu_{s}\rho}\int{\frac{d^{d}k}{(2 \pi)^d}}\frac{\partial}{\partial k_{\sigma}}\frac{k^{\nu_{1}}\cdots k^{\nu_{s-1}}}{\left[k^2-\mu^2\right]^{L}}+2L\int{\frac{d^{d}k}{(2 \pi)^d}}\frac{\partial}{\partial k_{\sigma}}\frac{k^{\nu_{1}}\cdots k^{\nu_{s}}k^{\rho}}{\left[k^2-\mu^2\right]^{L+1}}\Bigg]=0.
\end{align}
\noindent
The relations above are equivalent to the statements
\begin{align}
&\Upsilon_{1}^{(1)\sigma\nu_{1}\cdots\nu{s}}=0,\nonumber\\ \Upsilon_{0}^{(1)\sigma\nu_{2}\cdots\nu_{s}}=0,\quad\Upsilon_{0}^{(1)\sigma\nu_{1}\cdots\nu_{j-1}\nu_{j+1}\cdots \nu_{s}}=0,&\quad\Upsilon_{0}^{(1)\sigma\nu_{1}\cdots \nu_{s-1}}=0\quad\mbox{and}\quad\Upsilon_{0}^{(1)\sigma\nu_{1}\cdots \nu_{s}\rho}=0.
\end{align}

Therefore, we conclude that momentum routing invariance is guaranteed for quadratically divergent graphs only if surface terms are systematically set to zero. This result can be generalized to graphs with any kind of divergence proving that momentum routing invariance is verified only if we set all surface terms to zero. Although our results are restricted to one-loop order in perturbation theory, a general proof for arbitrary Feynman diagrams can also be developed, and we leave it to appendix \ref{proof}.

\subsection{Parametrizing Surface Terms}

Once we have demonstrated that the condition required to preserve momentum routing invariance is to set surface terms to zero, we
present a general
parametrization of divergent loop integrals, and through algebraic manipulations only we show how
the surface terms defined by equations (\ref{identsurface1}) and (\ref{identsurface}) can be shown to vanish. We begin by exemplifying
the logarithmic and quadratic surface terms to one-loop order. Taking the derivative of $I_{log}^{(1)}(\lambda^2)$ and
$I_{log}^{(1)\,\mu\nu}(\lambda^2)$ w.r.t $\lambda^2$, which must be satisfied by any regularization scheme, yields
\begin{align}
\frac{d I_{log}^{(1)}(\lambda^2)}{d \lambda^{2}}&=-\frac{b_{d}}{\lambda^{2}},\nonumber\\
\frac{d I_{log}^{(1)\,\mu\nu}(\lambda^2)}{d \lambda^{2}}&=-\frac{g_{\mu\nu}}{d}\frac{b_{d}}{\lambda^{2}}.
\end{align}

A general parametrization which obeys the relations above is given by
\begin{align}
I_{log}^{(1)}(\lambda^2)&=-\Bigg[a_{1}-b_{d}\ln\left(\frac{\Lambda^2}{\lambda^2}\right)\Bigg],\nonumber\\
I_{log}^{(1)\,\mu\nu}(\lambda^2)&=-\frac{g_{\mu\nu}}{d}\Bigg[a'_{1}-b_{d}\ln\left(\frac{\Lambda^2}{\lambda^2}\right)\Bigg],
\label{resilog}
\end{align}
\noindent
where $a_{1}$, $a'_{1}$ are arbitrary constants, $\Lambda$ is an ultraviolet cut-off, and $b_{d}$ is defined in eq. (\ref{bd}).

Introducing these parametrizations in the definition of $\Upsilon_{0}^{(1)\mu\nu}$
\begin{align}
\Upsilon_{0}^{(1)\mu\nu}=\int{\frac{d^{d}k}{(2 \pi)^d}}\frac{\partial}{\partial k_{\mu}}\frac{k^{\nu}}{(k^{2}-\lambda^{2})^{\frac{d}{2}}}=d\Bigg[\frac{g_{\mu\nu}}{d}I_{log}^{(1)}(\lambda^2)-I_{log}^{(1)\,\mu\nu}(\lambda^2)\Bigg],
\end{align}
\noindent
one finds it is zero only if $a_{1}=a'_{1}$, should MRI be respected.

A similar conclusion can be achieved in the case of quadratically divergences whose BDI's are given by $I_{quad}^{(1)}(\lambda^2)$, and $I_{quad}^{(1)\,\mu\nu}(\lambda^2)$. Similarly
\begin{align}
\frac{d I_{quad}^{(1)}(\lambda^2)}{d \lambda^{2}}&=\frac{(d-2)}{2}I_{log}^{(1)}(\lambda^2),\nonumber\\
\frac{d I_{quad}^{(1)\,\mu\nu}(\lambda^2)}{d \lambda^{2}}&=(d)I_{log}^{(1)\,\mu\nu}(\lambda^2).
\end{align}

Again, a parametrization that complies with the relations above is
\begin{align}
I_{quad}^{(1)}(\lambda^2)&=\frac{(d-2)}{2}\Bigg[c_{1}\Lambda^{2}+b_{d}\lambda^{2}\ln\left(\frac{\Lambda^2}{\lambda^2}\right)-(a_{1}-b_{d})\lambda^{2}\Bigg],\nonumber\\
I_{quad}^{(1)\,\mu\nu}(\lambda^2)&=\frac{g_{\mu\nu}}{2}\Bigg[c'_{1}\Lambda^{2}+b_{d}\lambda^{2}\ln\left(\frac{\Lambda^2}{\lambda^2}\right)-(a_{1}-b_{d})\lambda^{2}\Bigg],
\label{resiquad}
\end{align}
\noindent
where $c_{1}$, $c'_{1}$ are arbitrary constants. We notice that using these parametrizations in the definition of $\Upsilon_{2}^{(1)\mu\nu}$
\begin{align}
\Upsilon_{2}^{(1)\mu\nu}=\int{\frac{d^{d}k}{(2 \pi)^d}}\frac{\partial}{\partial k_{\mu}}\frac{k^{\nu}}{(k^{2}-\lambda^{2})^{\frac{d-2}{2}}}=(d-2)\Bigg[\frac{g_{\mu\nu}}{(d-2)}I_{quad}^{(1)}(\lambda^2)-I_{quad}^{(1)\,\mu\nu}(\lambda^2)\Bigg],
\end{align}
\noindent
the surface term is null provided $c_{1}=c'_{1}$.

To conclude this section we present the result for logarithmic divergences to $n$-loop order whose derivatives w.r.t $\lambda^2$ are expressed by
\begin{align}
\frac{d I_{log}^{(n)}(\lambda^2)}{d \lambda^{2}}&=-\frac{(n-1)}{\lambda^{2}}I_{log}^{(n-1)}(\lambda^2)+\frac{b_{d}}{\lambda^{2}}A^{(n)},\nonumber\\
\frac{d I_{log}^{(n)\,\mu\nu}(\lambda^2)}{d \lambda^{2}}&=-\frac{(n-1)}{\lambda^{2}}I_{log}^{(n-1)\,\mu\nu}(\lambda^2)+\frac{g_{\mu\nu}}{2}\frac{b_{d}}{\lambda^{2}}B^{(n)}.
\label{gerder}
\end{align}
\noindent
After some algebra one can demonstrate that the parametrization below respects (\ref{gerder}),
\begin{align}
I_{log}^{(n)}(\lambda^2)&=\sum\limits_{i=1}^{n}A^{(i)}\frac{(n-1)!}{(i-1)!}\!\Bigg[\!\frac{(-b_{d})}{(n-i+1)!}\ln^{n-i+1}\!\!\left(\frac{\Lambda^{2}}{\lambda^{2}}\right)+\sum\limits_{j=0}^{n-i}\frac{a_{n-j-i+1}}{j!(n-j-i)!}\ln^{j}\!\!\left(\frac{\Lambda^{2}}{\lambda^{2}}\right)\!\!\Bigg],\nonumber\\
I_{log}^{(n)\,\mu\nu}(\lambda^2)&=\frac{g_{\mu\nu}}{2}\!\sum\limits_{i=1}^{n}B^{(i)}\frac{(n-1)!}{(i-1)!}\!\Bigg[\!\frac{(-b_{d})}{(n-i+1)!}\ln^{n-i+1}\!\!\left(\frac{\Lambda^{2}}{\lambda^{2}}\right)+\sum\limits_{j=0}^{n-i}\frac{a'_{n-j-i+1}}{j!(n-j-i)!}\ln^{j}\!\!\left(\frac{\Lambda^{2}}{\lambda^{2}}\right)\!\!\Bigg],
\end{align}
\begin{align}
\mbox{where}\quad A^{(i)}\equiv\Gamma(d/2)\lim_{\delta\rightarrow0}\Bigg[&-(n-1)\sum\limits_{l=0}^{n-2}\binom{n-2}{l}\frac{(-1)^{1+l}}{\delta^{n-2}}\frac{\Gamma(1-\delta(n-2-l))}{\Gamma(d/2+1-\delta(n-2-l))}+\quad\quad\quad\quad\quad\quad\quad\quad\quad\nonumber\\&+\left(\frac{d}{2}\right)\sum\limits_{l=0}^{n-1}\binom{n-1}{l}\frac{(-1)^{1+l}}{\delta^{n-1}}\frac{\Gamma(1-\delta(n-1-l))}{\Gamma(d/2+1-\delta(n-1-l))}\Bigg],\nonumber\\
B^{(i)}\equiv\Gamma(d/2)\lim_{\delta\rightarrow0}\Bigg[&-(n-1)\sum\limits_{l=0}^{n-2}\binom{n-2}{l}\frac{(-1)^{1+l}}{\delta^{n-2}}\frac{\Gamma(1-\delta(n-2-l))}{\Gamma(d/2+2-\delta(n-2-l))}+\nonumber\\&+\left(\frac{d+2}{2}\right)\sum\limits_{l=0}^{n-1}\binom{n-1}{l}\frac{(-1)^{1+l}}{\delta^{n-1}}\frac{\Gamma(1-\delta(n-1-l))}{\Gamma(d/2+2-\delta(n-1-l))}\Bigg],
\end{align}
\noindent
and $a_{i}$, $a'_{i}$ are arbitrary constants. Once again, the surface terms to all orders
\begin{align}
\frac{1}{2}\sum_{j=1}^{n}\left(\frac{2}{d}\right)^j\frac{(n-1)!}{(n-j)!}\Upsilon_{0}^{(n)\mu\nu}=\frac{g^{\mu \nu}}{2}\sum_{j=1}^{n}\left(\frac{2}{d}\right)^j\!\frac{(n-1)!}{(n-j)!}I_{log}^{(l-j+1)}(\lambda^2)-I_{log}^{(n)\,\mu
\nu}(\lambda^2)=0,
\end{align}
\noindent
are zero only if $a_{i}=a'_{i}$.

Some comments are now in order. Firstly DReg automatically evaluates the coefficients $a_{1}=a'_{i}$ in (\ref{resilog}) and
 $c_{1}=c'_{i}=0$ in (\ref{resiquad}). Other regularization schemes will assign different values for such arbitrary coefficients
 which clearly demonstrates that they may violate MRI.

\section{Momentum Routing Invariance and QED Ward Identities}
\label{secqed}

In the following we study the consequences of imposing MRI by setting surface terms to zero. Our starting point is a diagrammatic proof of abelian gauge invariance to arbitrary order in perturbation theory. As is well known, such proof lies on the possibility of performing shifts in the internal momenta of divergent integrals \cite{Peskin}. Here we show that this is ultimately connected to MRI in the Feynman diagrams. To this end we use IReg in the integrals needed in the diagrammatic proof, and study under which conditions the Ward Identities are respected. In what follows we use a pictoric representation (figure \ref{figabc}) of the Ward Identities as in \cite{Peskin}, namely:

\begin{figure}[ht!]
\begin{center}
\includegraphics[scale=0.8]{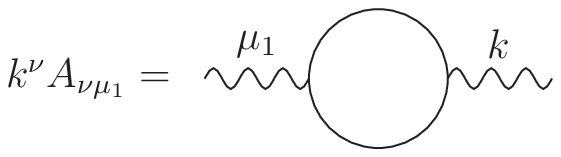}
\end{center}
\end{figure}
\begin{figure}[ht!]
\begin{center}
\includegraphics[scale=0.8]{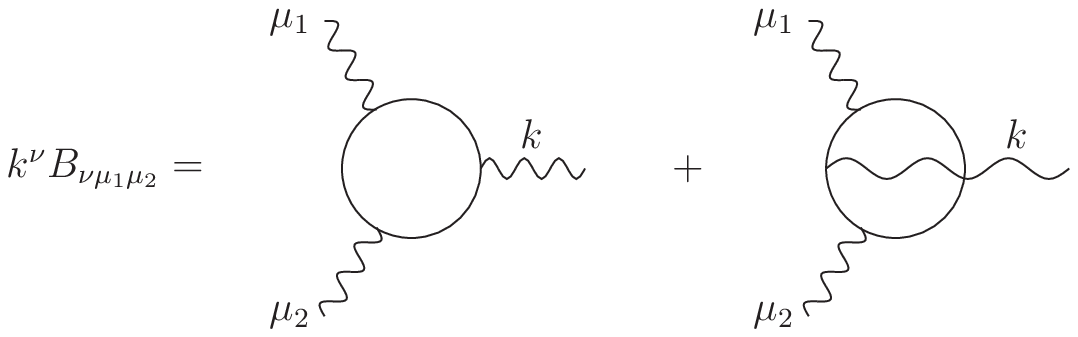}
\end{center}
\end{figure}
\begin{figure}[ht!]
\begin{center}
\includegraphics[scale=0.8]{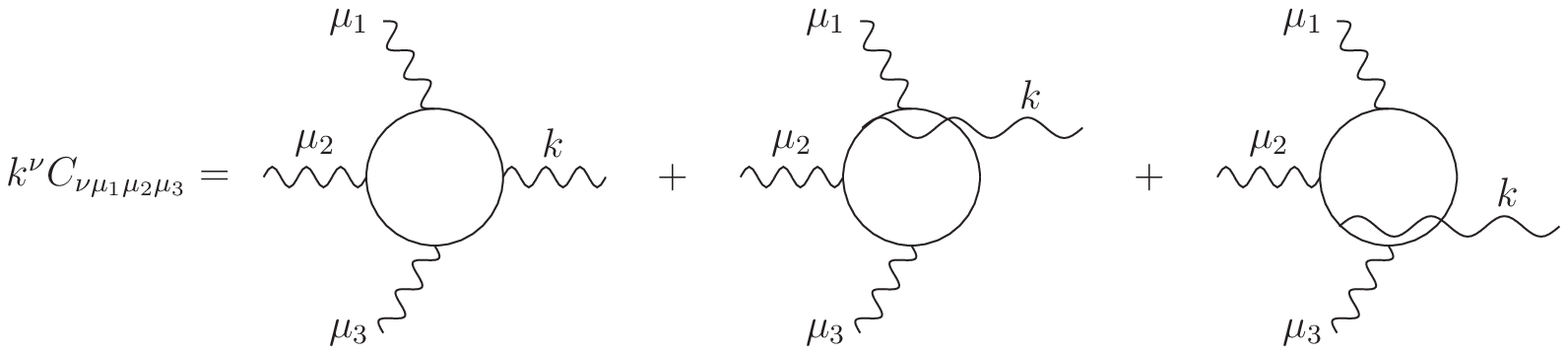}
\end{center}
\caption{\label{figabc} Pictoric representation of QED Ward identities $k^{\nu}A_{\nu\mu_{1}}=0$, $k^{\nu}B_{\nu\mu_{1}\mu_{2}}=0$, and $k^{\nu}C_{\nu\mu_{1}\mu_{2}\mu_{3}}=0$. Diagrams with more than four external legs are finite and shifts are obviously allowed.}
\end{figure}
\noindent
Explicitly,
\begin{align}
k^{\nu}A_{\nu\mu_{1}}&=Tr\int_{p}\gamma_{\mu_1}\left(\frac{1}{\slashed p +\slashed \alpha +\slashed k}\right)\slashed k\left(\frac{1}{\slashed p +\slashed \alpha}\right),\\
k^{\nu}B_{\nu\mu_{1}\mu_{2}}&=Tr\int_{p}\gamma_{\mu_2}\left(\frac{1}{\slashed p +\slashed \beta +\slashed k +\slashed q}\right)\gamma_{\mu_1}\left(\frac{1}{\slashed p +\slashed \beta +\slashed k}\right)\slashed k\left(\frac{1}{\slashed p +\slashed \beta}\right)\nonumber\\&+Tr\int_{p}\gamma_{\mu_2}\left(\frac{1}{\slashed p +\slashed \beta +\slashed k +\slashed q}\right)\slashed k\left(\frac{1}{\slashed p +\slashed \beta +\slashed q}\right)\gamma_{\mu_1}\left(\frac{1}{\slashed p +\slashed \beta}\right),\\
k^{\nu}C_{\nu\mu_{1}\mu_{2}\mu_{3}}&=Tr\int_{p}\gamma_{\mu_3}\!\left(\frac{1}{\slashed p +\slashed \delta +\slashed k +\slashed Q}\right)\!\gamma_{\mu_2}\!\left(\frac{1}{\slashed p +\slashed \delta +\slashed k +\slashed q_{1}}\right)\!\gamma_{\mu_1}\!\left(\frac{1}{\slashed p +\slashed \delta +\slashed k}\right)\!\slashed k\!\left(\frac{1}{\slashed p +\slashed \delta}\right)\!\nonumber\\
&+Tr\int_{p}\gamma_{\mu_3}\!\left(\frac{1}{\slashed p +\slashed \delta +\slashed k +\slashed Q}\right)\!\gamma_{\mu_2}\!\left(\frac{1}{\slashed p +\slashed \delta +\slashed k +\slashed q_{1}}\right)\!\slashed k\!\left(\frac{1}{\slashed p +\slashed \delta +\slashed q_{1}}\right)\!\gamma_{\mu_1}\!\left(\frac{1}{\slashed p +\slashed \delta}\right)\!\nonumber\\
&+Tr\int_{p}\gamma_{\mu_3}\!\left(\frac{1}{\slashed p +\slashed \delta +\slashed k +\slashed Q}\right)\!\slashed k\!\left(\frac{1}{\slashed p +\slashed \delta +\slashed Q}\right)\!\gamma_{\mu_2}\!\left(\frac{1}{\slashed p +\slashed \delta +\slashed q_{1}}\right)\!\gamma_{\mu_1}\!\left(\frac{1}{\slashed p +\slashed \delta}\right)\!,
\end{align}
\noindent
where $\alpha$, $\beta$, and $\delta$ are arbitrary routings and $\slashed Q\equiv\slashed q_{1} +\slashed q_{2}$. By using IReg we finally obtain:
\begin{align}
k^{\nu}A_{\nu\mu_{1}}&=-4\Gamma_{2}^{(1,2)}k_{\mu_1}+4\left(\Gamma_{0}^{(1,2)}-4\Gamma_{0}^{(1,4)}\right)\big[k\cdot(k+2\alpha)\alpha_{\mu_{1}}+(k+\alpha)^{2}k_{\mu_{1}}\big],\\
k^{\nu}B_{\nu\mu_{1}\mu_{2}}&=-4\left(\Gamma_{0}^{(1,2)}-4\Gamma_{0}^{(1,4)}\right)\big[g_{\mu_{1}\mu_{2}}k\cdot(k+q+2\beta)+k_{\mu_{1}}(k+q+2\beta)_{\mu_{2}}+\nonumber\\&\quad\quad\quad\quad\quad\quad\quad\quad\quad\quad\quad+k_{\mu_{2}}(k+q+2\beta)_{\mu_{1}}\big],\\
k^{\nu}C_{\nu\mu_{1}\mu_{2}\mu_{3}}&=4\left(\Gamma_{0}^{(1,2)}-4\Gamma_{0}^{(1,4)}\right)\big[g_{\mu_{1}\mu_{2}}k_{\mu_{3}}+g_{\mu_{1}\mu_{3}}k_{\mu_{2}}+g_{\mu_{2}\mu_{3}}k_{\mu_{1}}\big].
\end{align}
\noindent
with surface terms defined in eq. (\ref{idets}). We notice that Ward identities are fulfilled provided
\be
\Gamma_{2}^{(1,2)}=\Gamma_{0}^{(1,2)}=\Gamma_{0}^{(1,4)}=0.
\ee
\noindent
Thus by adopting an abelian gauge invariant regularization (setting surface terms to zero) one automatically preserves MRI. 

Conversely, we could study the conditions under which the diagrams involved in the diagrammatic proof \cite{Peskin} of gauge invariance respect MRI
\begin{figure}[ht!]
\begin{center}
\includegraphics[scale=0.8]{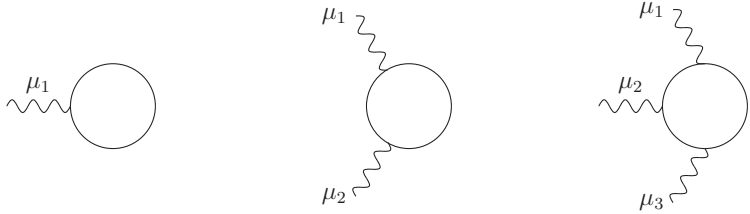}
\vspace{-0.9cm}
\end{center}
\caption{ Diagrams needed in the diagramatic proof of abelian gauge invariance.}
\end{figure}
\vspace{-0.1cm}

Performing the calculation one notices that MRI is respected if 
\be
\Gamma_{2}^{(1,2)}=\Gamma_{0}^{(1,2)}=\Gamma_{0}^{(1,4)}=0,
\ee
\noindent
which are the \textit{same} conditions to preserve the Ward identities. Therefore we conclude that MRI is a necessary and sufficient condition to preserve abelian gauge symmetry at arbitrary loop order. This should be emphasized that the previous conclusion regards a theory free of chiral couplings (proportional to $\gamma_{5}$). This feature can be easily seen in the Adler-Bardeen-Bell-Jackiw (ABJ) chiral anomaly in which the vector Ward identity must be violated if momentum routing invariance is to be respected. 

\begin{figure}[ht!]
\begin{center}
\includegraphics[scale=0.9]{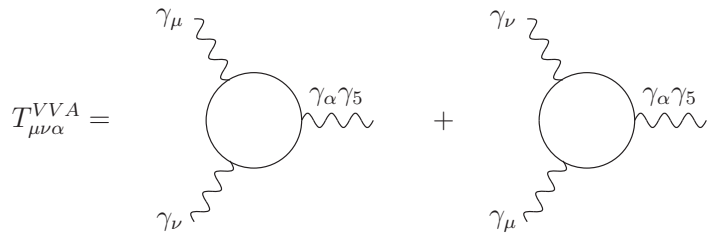}
\end{center}
\caption{\label{figtavv} Diagrams of the Adler-Bardeen-Bell-Jackiw chiral anomaly with an arbitrary momentum routing.}
\end{figure}

The ABJ chiral anomaly was already studied using IReg \cite{Prd2}. The relevant diagrams are depicted in figure \ref{figtavv}, whose amplitudes are given by
\begin{align}
T_{\mu \nu \alpha}^{VVA}=&-Tr\int_{p}\gamma_{\nu}\left(\frac{1}{\slashed p +\slashed \delta +\slashed k }\right)\gamma_{\mu}\left(\frac{1}{\slashed p +\slashed \delta}\right)\gamma_{\alpha}\gamma_{5}\left(\frac{1}{\slashed p +\slashed \delta+ \slashed k +\slashed q}\right)\nonumber\\&-Tr\int_{p}\gamma_{\mu}\left(\frac{1}{\slashed p +\slashed \lambda +\slashed q}\right)\gamma_{\nu}\left(\frac{1}{\slashed p +\slashed \lambda}\right)\gamma_{\alpha}\gamma_{5}\left(\frac{1}{\slashed p +\slashed \lambda + \slashed k + \slashed q}\right),
\end{align}
\noindent
where $\delta$ and $\lambda$ are arbitrary momentum routings.

In order to make the discussion that follows clearer, we rewrite $\delta - \lambda = (\delta'-\lambda')(q-k)$ where $\delta'$ and $\lambda'$ are arbitrary (dimensionless) constants. Using now IReg \cite{Prd2} one arrives at the vector and axial Ward identities\footnote{A comment is in order: textbook calculations using Dimensional Regularization \cite{Bertlmann} are usually done by firstly inserting the external momentum into the amplitude before integration. This operation does not commute with integrating the amplitude and after that contracting with the external momenta. While the former allows the amplitude to be written as a surface term of the form $\int_p [f(p+k)-f(p)]_{\nu \alpha}$, the latter does not. This is the origin of the term $\frac 1{4\pi^2}$ in the vector Ward identity.}, which cannot be satisfied simultaneously
\begin{align}
k^{\mu}T_{\mu \nu \alpha}^{VVA}&= \left\{-\frac 1{4\pi^2} - 4i(\delta'-\lambda'+1)\Gamma_{0}^{(1,2)}
\right\} \epsilon_{\mu \nu \alpha \beta}k^{\mu}q^{\beta} \nonumber \\
q^{\nu}T_{\mu \nu \alpha}^{VVA}&=\left\{\frac 1{4\pi^2} + 4i(\delta'-\lambda'+1)\Gamma_{0}^{(1,2)}
\right\}\epsilon_{\mu \nu \alpha \beta}q^{\nu}k^{\beta} \nonumber \\
(k+q)^{\alpha}T_{\mu \nu \alpha}^{VVA}&=-8i(\delta'-\lambda'+1)\Gamma_{0}^{(1,2)} \epsilon_{\mu \nu \alpha \beta}k^{\alpha}q^{\beta},
\label{wi}
\end{align}
\noindent

A comment is in order to clarify the interplay between surface terms and MRI in the presence of anomalies. We notice that by setting $\Gamma_{0}^{(1,2)}$ to zero one implements MRI since the Ward identities will be independent of $\delta'$ and $\lambda'$. However, this choice for $\Gamma_{0}^{(1,2)}$ strikingly spoils the democracy that the calculation scheme should preserve between the vector and axial Ward identities which must be fixed on physical grounds \cite{Jackiw:1999qq}. In other words, the surface term should be left arbitrary and a new constraint should be imposed on the theory. In the case of the ABJ chiral anomaly, the pion decay into two photons (experimental data) requires the conservation of the vector current. Therefore, the surface term is required to assume a non-vanishing value such that
\be
4i(\delta'-\lambda'+1)\Gamma_{0}^{(1,2)}=-\frac 1{4\pi^2},
\ee
\noindent
which reflects the breaking of MRI.

\section{Examples}
\label{secex}

\subsection{Momentum Routing Invariance and Supersymmetry}

In this section we investigate the connection between MRI and supersymmetry preservation in perturbation theory.
For this purpose we study the massless Wess-Zumino model in component field formalism up to two-loops.

\begin{figure}[ht!]
\begin{center}
\includegraphics[scale=0.8]{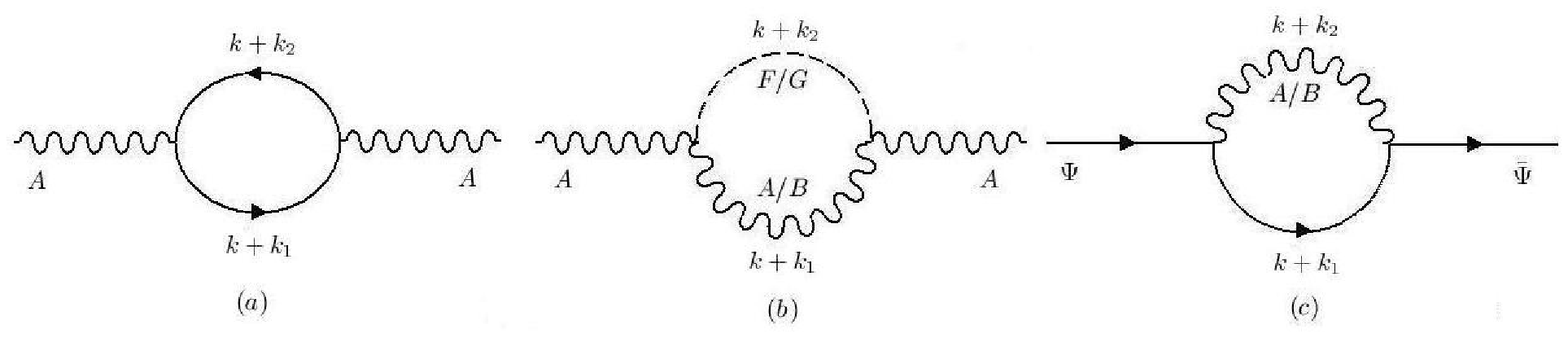}
\end{center}
\vspace{-1.0cm}
\caption{\label{fig1lwz} One-loop diagrams with most generic routing of internal lines. In the figure, $k$ is the
internal momentum and $p \equiv k_{1} - k_{2}$ is the external momentum.}
\end{figure}

The Lagrangian\footnote{The Feynman rules can be found in Appendix \ref{appwz}} for the model is \cite{wess} given by:
\begin{align}
\mathcal{L} = &- \frac{1}{2} \big[ (\partial_{\mu} A)^{2} + (\partial_{\mu} B)^{2} + \bar{\Psi}\gamma_{\mu}\partial^{\mu}\Psi
-F^{2} -G^{2} \big]-
\nonumber
\\
& - g[F(A^2 -B^2) - 2GAB - \bar{\Psi}(A + i\gamma_{5}B )\Psi] ,
\label{lwz}
\end{align}
where $A$ and $B$ are bosonic (scalar and pseudoscalar, respectively), fields. The field $\Psi$ is a fermionic field of Majorana
 type $\Psi_{M}^{\chi}$, and $F$ and $G$, auxiliary fields. In what follows we investigate the Ward identity
 for two-point functions of the $A$ and $\Psi$ fields
\be
\Gamma_{AA}(p) + i\pbruto \Gamma_{\Psi\bar{\Psi}}(p) = 0 .
\label{idwt}
\ee
To one-loop order the relevant two-point functions are depicted in figure \ref{fig1lwz}.
\vspace{-0.5cm}

\noindent
Note the arbitrary routing of the internal lines which complies with energy-momentum conservation at the
vertices in its most generic form. The two-point function reads:
\begin{align}
\Gamma_{AA}^{(a)} = 4g^{2} \bigg\{& 2I_{quad}^{(1)}(\lambda^{2}) - (\kum - \kdois)^{2}I_{log}^{(1)}(\lambda^{2}) -
[\kummu \kumnu + \kdoismu \kdoisnu] \Upsilon_{0}^{(1)\mu\nu} -
\nonumber \\
& - (\kum - \kdois)^{2}\bigg[ 2b_{4} - b_{4}\lnkumdois \bigg] \bigg\} ,
\label{gaa1f}
\end{align}
\noindent
where $b_{4}\equiv\frac{i}{(4\pi)^{2}}$.

Notice that physical amplitudes must depend solely on the combination $\kum - \kdois=p$.
However in equation (\ref{gaa1f}) we see immediately an explicit dependence on the particular choice of internal momenta, which,
 however is multiplied by a surface term. Likewise
\begin{align}
\Gamma_{AA}^{(b)} & = -8g^{2} \big\{ \Iquad - \kummu \kumnu \Upsilon_{0}^{(1)\mu\nu} \big\},
\label{gaa2f}\\
\Gamma_{\Psi\bar{\Psi}}^{(c)} & = -8ig^2 \bigg\{ \frac{1}{2}(\kumbruto - \kdoisbruto)\Ilog +
\frac{1}{2}\gamma_{\mu}(\kumnu + \kdoisnu)\Upsilon_{0}^{(1)\mu\nu} +
\nonumber \\
&\ \ \ \ \ \ \ \ \ \ \ \ \ + \frac{1}{2}(\kumbruto - \kdoisbruto)\bigg[ 2b_{4} -b_{4}\lnkumdois \bigg]\bigg\}.
\label{gxxf}
\end{align}

Now, substituting the above expressions in Ward identity (\ref{idwt}), we obtain:
\begin{align}
\Gamma_{AA}(p) + i\pbruto \Gamma_{\Psi\bar{\Psi}}(p) &=
4g^{2} \big\{  \kummu \kumnu - \kdoismu \kdoisnu +
(\kumbruto - \kdoisbruto)\gamma_{\mu}(\kumnu + \kdoisnu) \big\} \Upsilon_{0}^{(1)\mu\nu} .
\end{align}
\noindent
In order to preserve this identity we must demand that
\be
\Upsilon_{0}^{(1)\mu\nu} = 0.
\ee
\noindent
To make contact with the literature \cite{MCKEON}, we rewrite $\kummu \rightarrow \alpha p_{\mu}$
and $\kdoismu \rightarrow (\alpha -1)p_{\mu}$, where $\alpha$ is an arbitrary constant. In terms of this parametrization
we get
\begin{align}
\Gamma_{AA}(p) + i\pbruto \Gamma_{\Psi\bar{\Psi}}(p) =
4g^{2} [ 2\alpha-1] \big\{ p_{\mu} p_{\nu} +
\pbruto \gamma_{\mu}p_{\nu}\big\} \Upsilon_{0}^{(1)\mu\nu} .
\end{align}
\noindent
Now we see that it is possible to satisfy the Ward identity by choosing a specific routing, i.e. $\alpha = \frac{1}{2}$. This
approach was discussed in \cite{MCKEON}. In our viewpoint, $\alpha$ should be left arbitrary and the surface term set to zero on MRI
grounds.

\begin{figure}[ht!]
\begin{center}
\includegraphics[scale=0.8]{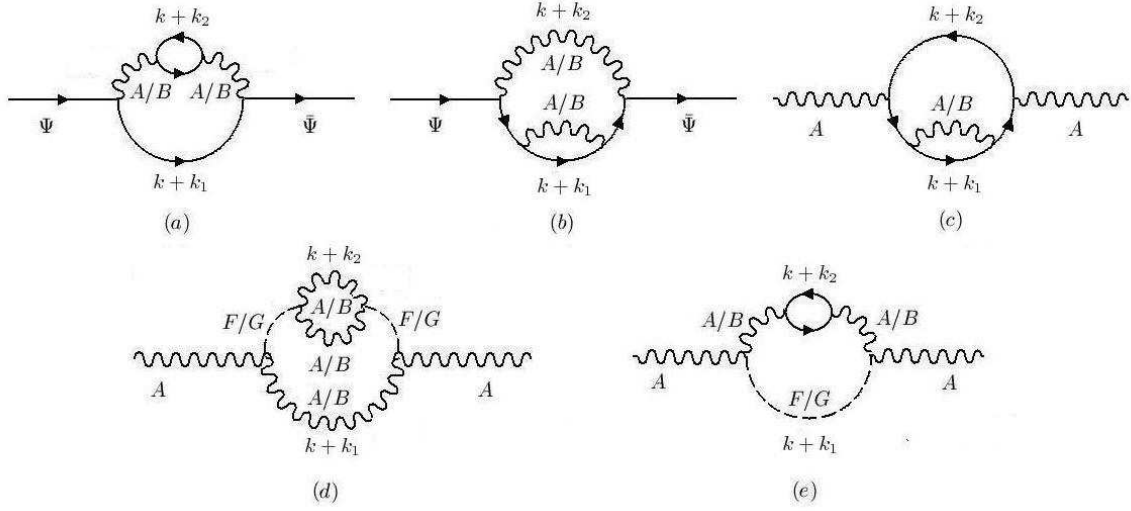}
\end{center}
\caption{\label{fig2lwz} Two-loop diagrams with most generic routing of internal lines. In the figure, $k$ is the
internal momentum and $p \equiv k_{1} - k_{2}$ is the external momentum.}
\end{figure}

We perform the same analysis at two-loop order. Depicted in figure \ref{fig2lwz} are the diagrams which contribute to the Ward identity, again with arbitrary momentum routing in the loops.

\vspace{0.5cm}
\noindent
The results for the two-point functions are
\begin{align}
\Gamma_{\Psi \bar{\Psi}}^{(a)} &= 16b_{4}g^{4} \bigg\{ (\kumbruto - \kdoisbruto)\bigg[ -\Ilogd + \frac{5}{2}\Ilog \bigg] -
 \gamma_{\mu} (\kumnu + \kdoisnu) \Upsilon_{0}^{(2)\mu\nu} \bigg\} + \mbox{finite} ,
\label{gxx_2l1}\\
\Gamma_{\Psi \bar{\Psi}}^{(b)} &= 16b_{4}g^{4} \bigg\{ (\kumbruto - \kdoisbruto)\bigg[ -\Ilogd + \frac{3}{2}\Ilog \bigg] -
\gamma_{\mu} (\kumnu + \kdoisnu) \Upsilon_{0}^{(2)\mu\nu} \bigg\} + \mbox{finite} ,
\label{gxx_2l2}\\
\Gamma_{AA}^{(c)} & = 32ib_{4}g^{4} \bigg\{ -2\Iquadd + 4\Iquad + (\kum - \kdois)^{2}\bigg[ \Ilogd - \frac{5}{2}\Ilog \bigg] +
\nonumber \\
& \ \ \ \ \ \ \ \ \ \ \ \ \ \ \ \ + \big[  (\kummu - \kdoismu)(\kumnu - \kdoisnu) + 2\kummu  (\kumnu + \kdoisnu) \big]\Upsilon_{0}^{(2)\mu\nu} \bigg\} + \mbox{finite} ,
\label{gaa_2l1}\\
\Gamma_{AA}^{(d)} & = -32ib_{4}g^{4} \bigg\{ -\Iquadd + 2\Iquad - \frac{1}{2}(\kum - \kdois)^{2}\Ilog +
\nonumber \\
& \ \ \ \ \ \ \ \ \ \ \ \ \ \ \ \ \ \ + \big[  (\kummu - \kdoismu)(\kumnu - \kdoisnu) + \kdoismu  (2\kumnu - \kdoisnu) \big]\Upsilon_{0}^{(2)\mu\nu} \bigg\} + \mbox{finite} ,
\label{gaa_2l2}\\
\Gamma_{AA}^{(e)} & = 32ib_{4}g^{4} \bigg\{ \Iquadd - 2\Iquad - \kdoismu\kdoisnu \Upsilon_{0}^{(2)\mu\nu} \bigg\} ,
\label{gaa_2l3}
\end{align}
\noindent
in which the finite part depends only on external momentum $p\equiv \kum - \kdois$.

With these results, Ward identity (\ref{idwt}) can be written as:
\begin{align}
\Gamma_{AA}(p) + i\pbruto \Gamma_{\Psi\bar{\Psi}}(p) = 32ibg^{4} \big\{ &2\kummu(\kumnu + \kdoisnu) + 2\kdoismu(\kumnu - \kdoisnu) -
\nonumber \\
& - (\kumbruto - \kdoisbruto)\gamma_{\mu}(\kumnu + \kdoisnu) \big\}\Upsilon_{0}^{(2)\mu\nu} ,
\end{align}
\noindent
where $\Upsilon_{0}^{(2)\mu\nu}$ is a surface term in second order perturbation theory. Here again the connection
between momentum routing invariance and supersymmetry breaking becomes apparent. The same breaking
mechanism found for gauge theories are at work also in supersymmetry. Once more, the way to both implement momentum
routing invariance and preserve the Ward identity is to set
\be
\Upsilon_{0}^{(2)\mu\nu} = 0.
\ee

\subsection{Scalar Theories}

Having demonstrated that the imposition of momentum routing invariance automatically preserves abelian gauge symmetry to all-orders, and conjectured that it may be also the root of supersymmetry preservation, we wonder which would be the consequences of MRI breaking in a theory with low symmetry content. We study massless $\phi_{6}^{3}$ theory by calculating its $\beta$-function at two-loop order \cite{Adriano}. The Feynman diagrams we will need are depicted in figure \ref{figphi}.

\begin{figure}[ht!]
\begin{center}
\includegraphics{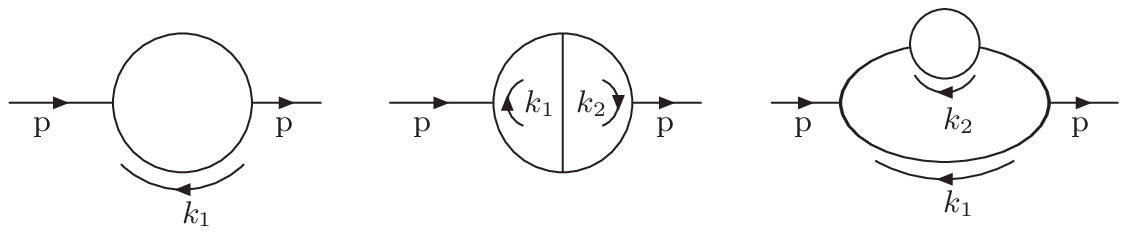}
\end{center}
\end{figure}
\vspace{-0.5cm}
\begin{figure}[ht!]
\begin{center}
\includegraphics{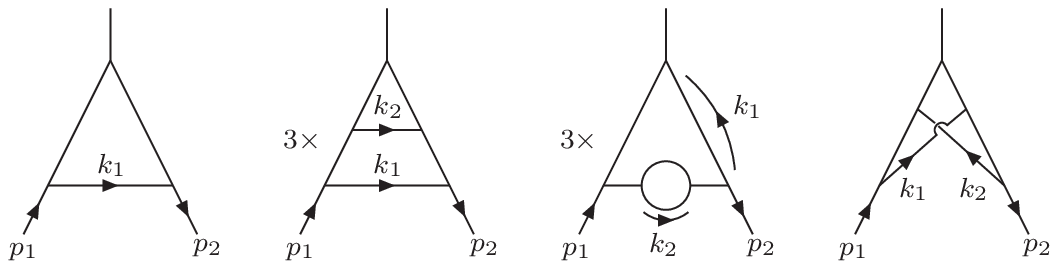}
\end{center}
\caption{\label{figphi} Diagrams contributing up to two-loop order of $\phi_{6}^{3}$ theory. $\Xi$ stands for two-point functions and $\Delta$ stands for three-point ones.}
\end{figure}

Since we want to stress the connection between surface terms and MRI, we will use an arbitrary momentum routing in the diagrams above. However, following the calculation outlined at \cite{Adriano}, one can easily see that only the momentum routing of the subgraph of the third and sixth diagrams will affect the coefficients of the $\beta$-function at two-loop order. Thus, we will use the following convention

\begin{figure}[ht!]
\begin{center}
\includegraphics{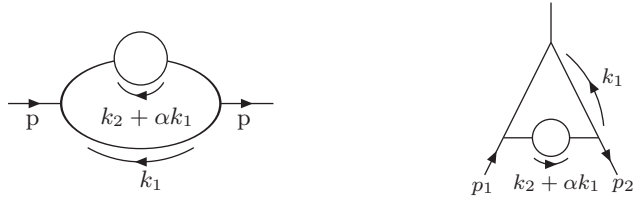}
\end{center}
\vspace{-0.3cm}
\caption{Diagrams whose choice of momentum routing may affect the coefficients of the $\beta$-function at two-loop order.}
\end{figure}

Adopting a minimal subtraction scheme which in IReg corresponds to subtracting BDI's only, one obtains the following two-loop corrections for propagator and vertex
\begin{align}
\Xi=&-g^{2}\Bigg[\frac{p^2}{6}I_{log}^{(1)}(\lambda^{2})\Bigg]+ig^{4}\Bigg\{\Bigg[\frac{5b_{6}}{18}I_{log}^{(2)}(\lambda^2)-\Bigg(\frac{29b_{6}}{54}-\frac{(3\alpha^2-3\alpha+1)\Gamma_{0}^{(1,2)}}{9}\Bigg)I_{log}^{(1)}(\lambda^2)\Bigg]p^2\Bigg\}\nonumber\\
&\quad\quad\quad+\mbox{terms that do not contribute to the $\beta$-function},\\
\Delta=&-g^3\Bigg[I_{log}^{(1)}(\lambda^{2})\Bigg]+ig^{5}\Bigg[\frac{5b_{6}}{2}I_{log}^{(2)}(\lambda^2)\!-\!\left(\frac{17b_{6}}{3}-(3\alpha^2-3\alpha+1)\Gamma_{0}^{(1,2)}\right)I_{log}^{(1)}(\lambda^2)\Bigg]\nonumber\\
&\quad\quad\quad+\mbox{terms that do not contribute to the $\beta$-function},
\end{align}
\noindent
\begin{equation*}
\mbox{where}\quad b_{6}\equiv-\frac{1}{2}\frac{i}{(4\pi)^{3}}. \quad  \quad \quad\quad\quad\quad\quad\quad\quad\quad\quad\quad\quad\quad\quad\quad\quad\quad\quad\quad\quad\quad\quad\quad\quad\quad\quad\quad\quad\quad
\end{equation*}

With these results, the $\beta$-function can be calculated yielding
\begin{align}
\beta=-\frac{3g^3}{4(4\pi)^{3}}-\frac{125g^5}{144(4\pi)^{6}}+\frac{ig^{5}\left[12(\alpha^2-\alpha)+5\right]\Gamma_{0}^{(1,2)}}{6(4\pi)^{3}}+O(g^{6}).
\end{align}

As is well known the coefficients of the $\beta$-function up to two-loop order in a mass independent scheme are universal. We notice that we obtain the universal values of the $\beta$-function only if we set $\Gamma_{0}^{(1,2)}=0$, which is the same condition to preserve MRI. On the other hand it becomes also clear from our expression for the beta function that should the surface term be non-vanishing the two loop universal beta function coefficient would be momentum routing dependent (in our case would depend on $\alpha$). Therefore, we conclude that even in a theory with poor symmetry content such as $\phi_{6}^{3}$, momentum routing invariance is important since it is responsible for the preservation of the universal values of the $\beta$-function.

\section{Concluding Remarks}
\label{secco}

 Momentum routing invariance (MRI) is clearly a symmetry of any Feynman graph, as allowed by energy-momentum conservation in the vertices. 
 However, divergencies which typically appear in loop calculations may simultaneously break MRI and important symmetries of the underlying
 model, such as gauge and supersymmetry. We demonstrated for abelian gauge theories that the offending terms are multiplied by surface terms which once set to zero automatically render the regularization invariant besides implementing MRI. We illustrated with examples that MRI is at the root of supersymmetry breaking as well. Therefore, we may conjecture that an invariant four dimensional regularization framework should comply with MRI. We derived the formal expressions of surface terms to arbitrary order in perturbation theory. In presence of anomalous processes we use the case of the  Adler-Bardeen-Bell-Jackiw (ABJ) chiral anomaly to illustrate that the same procedure can be applied to isolate the anomaly specific MRI violating terms, which can be fixed either to conserve the vector or the axial-vector current, in a manifestely ``democratic'' way. For theories with poor symmetry content, MRI manisfests itself as an important ingredient in the calculation of the universal coefficients of the $\beta$-function. In recent works we have verified that MRI is also important to preserve the Slavnov Taylor identities for non-abelian gauge theories and supersymmetric gauge theories \cite{BETAYM}, \cite{David}, \cite{Eloy2}, \cite{Eloy}. A formal proof for these cases is an ongoing work based on diagrammatics and the quantum action principle \cite{quantumaction}.
 
 Any regularization scheme should comply with MRI. We propose Implicit Regularization as an ideal framework to implement such program in the physical dimension of the quantum field theoretical model, applicable to models where dimensional methods may fail.
 
\acknowledgments

L. C. Ferreira, A. L. Cherchiglia and Marcos Sampaio acknowledge financial support by CNPq.  A. L. Cherchiglia and Marcos Sampaio acknowledge financial support by FAPEMIG. Partially supported by Funda\c{c}\~ao para a Ci\^encia e Tecnologia, Project CERN/FP116334/2010.

\appendix
\section{$N$-loop order proof connecting Momentum Routing Invariance and Surface Terms}
\label{proof}

We present in the following the demonstration for two-loop graphs and state how it can be further generalized to an arbitrary number of loops.

Assume that the amplitude of a two-loop process is given by,
\be
A^{(2)}\equiv\int\frac{d^{d}k_{1}}{(2 \pi)^d}\frac{d^{d}k_{2}}{(2 \pi)^d}f(k_{1}+\alpha_{1},k_{2}+\alpha_{2},q_{i}),
\ee
\noindent
in which space-time and internal group algebra have already been performed, and $\alpha_{1}$, $\alpha_{2}$ are arbitrary momentum routings that depend only on the external momenta $q_{i}$. Once again, momentum routing invariance is guaranteed by the condition
\be
\int\frac{d^{d}k_{1}}{(2 \pi)^d}\frac{d^{d}k_{2}}{(2 \pi)^d}\left[\prod\limits_{j=1}^{2}exp\left(\alpha_{j}^{\sigma_{j}}\frac{\partial}{\partial k_{j}^{\sigma_{j}}}\right)-\prod\limits_{j=1}^{2}exp\left(\beta_{j}^{\sigma_{j}}\frac{\partial}{\partial k_{j}^{\sigma_{j}}}\right)\right]f(k_{1},k_{2},q_{i})=0.
\label{condition2}
\ee

At this point we must use the rules of IReg in one of the integrals, however, which one must be evaluated first is not clear \textit{a priori}. Solving this problem was the main purpose of \cite{Adriano} in which we presented a prescription that systematizes the order of integration for  multi-loop Feynman diagrams. Therefore, using this prescription, amplitude $A^{(2)}$ can be decomposed in three cases:
\be
A^{(2)}=A_{k_{1}}+A_{k_{2}}+A_{fin},
\ee
\noindent
where in $A_{k_{i}}$ the integration over $k_{i}$ must be performed first, and $A_{fin}$ contains only finite terms which do not contribute. Once the order of integration has been determined we notice that condition (\ref{condition2}) reduces to
\be
\int\frac{d^{d}k_{1}}{(2 \pi)^d}\frac{d^{d}k_{2}}{(2 \pi)^d}\prod\limits_{j=1}^{2}\left(\frac{\partial}{\partial k_{j}^{\sigma_{j}}}\right)^{m_{j}}\left[A_{k_{1}}+A_{k_{2}}\right]=0,\quad\forall\; m_{j}\in\mathbb{N}.
\ee
Since the proof for $A_{k_{1}}$ is essentially the same for $A_{k_{2}}$, we just consider the latter in the following. Remembering that $A_{k_{2}}$ is a function of $k_{1}$, $k_{2}$, and $q_{i}$ we notice that it can be rewritten as
\be
A_{k_{2}}=\frac{1}{\prod\limits_{r}\left[(k_{1}+l_{r}(q_{i}))^2-\mu^2\right]}\frac{g(k_{2},k_{1},q_{i})}{\prod\limits_{j}\left[(k_{2}+l_{j}(k_{1},q_{i}))^2-\mu^2\right]}.
\ee

Therefore, we just have to prove that the condition below always holds
\begin{align}
\int\frac{d^{d}k_{1}}{(2 \pi)^d}\left(\frac{\partial}{\partial k_{1}^{\sigma_{1}}}\right)^{m_{1}}\!\!\!\!\!\frac{1}{\prod\limits_{r}\left[(k_{1}+l_{r}(q_{i}))^2-\mu^2\right]}\int\frac{d^{d}k_{2}}{(2 \pi)^d}\left(\frac{\partial}{\partial k_{2}^{\sigma_{2}}}\right)^{m_{2}}\!\!\!\!\!\frac{g(k_{2},k_{1},q_{i})}{\prod\limits_{j}\left[(k_{2}+l_{j}(k_{1},q_{i}))^2-\mu^2\right]}=0.
\end{align}

We have two cases: $m_{2}\neq0$, and $m_{2}=0$. In the first one we can use the one-loop results to obtain terms of the type
\be
\int\frac{d^{d}k_{1}}{(2 \pi)^d}\left(\frac{\partial}{\partial k_{1}^{\sigma_{1}}}\right)^{m_{1}}\frac{1}{\prod\limits_{r}\left[(k_{1}+l_{r}(q_{i}))^2-\mu^2\right]}\times h_{\nu_{1}\cdots\nu_{s}}(k_{1},q_{i})\Upsilon_{j}^{(1)\nu_{1}\cdots\nu_{s}}=0
\ee
\noindent
which is satisfied if all one-loop order surface terms are systematically set to zero. In the second case, we must use the rules of IReg in the $k_{2}$ integral to obtain
\begin{align}
\int\frac{d^{d}k_{1}}{(2 \pi)^d}\left(\frac{\partial}{\partial k_{1}^{\sigma_{j}}}\right)^{m_{1}}\frac{1}{\prod\limits_{r}\left[(k_{1}+l_{r}(q_{i}))^2-\mu^2\right]}\times h(k_{1},q_{i})\left[\mbox{BDI's} + \mbox{one-loop surface terms}\right]+\nonumber\\+\int\frac{d^{d}k_{1}}{(2 \pi)^d}\left(\frac{\partial}{\partial k_{1}^{\sigma_{j}}}\right)^{m_{1}}\frac{1}{\prod\limits_{r}\left[(k_{1}+l_{r}(q_{i}))^2-\mu^2\right]}\times\left[ \sum\limits_{s=1}^{2}a_{s}(k_{1},q_{i})\ln^{s-1}(k_{1},q_{i})\right]=0
\end{align}
\noindent
where $a_{s}(k_{1},q_{i})$ is a polynomial. Since $m_{1}\neq0$ we may use a similar analysis done in the one-loop case to show that the first integral is proportional to one-loop surface terms whereas the second is proportional to one-loop and two-loop ones. Therefore, we achieve our major goal: all terms involved in (\ref{condition2}) are proportional to surface terms of one-loop and two-loop order, showing that momentum routing is guaranteed only if surface terms are systematically set to zero. This conclusion is not restricted to two-loop oder, since a similar demonstration can be performed to graphs with an arbitrary number of loops. Thus, we may state that the condition to implement momentum routing invariance is to set surface terms of all orders to zero.

\section{Wess-Zumino Feynman rules}
\label{appwz}

The Feynman rules for the massless Wess-Zumino model are:

\vspace{4.0cm}
\begin{figure}[ht!]
\begin{center}
\includegraphics[scale=0.8]{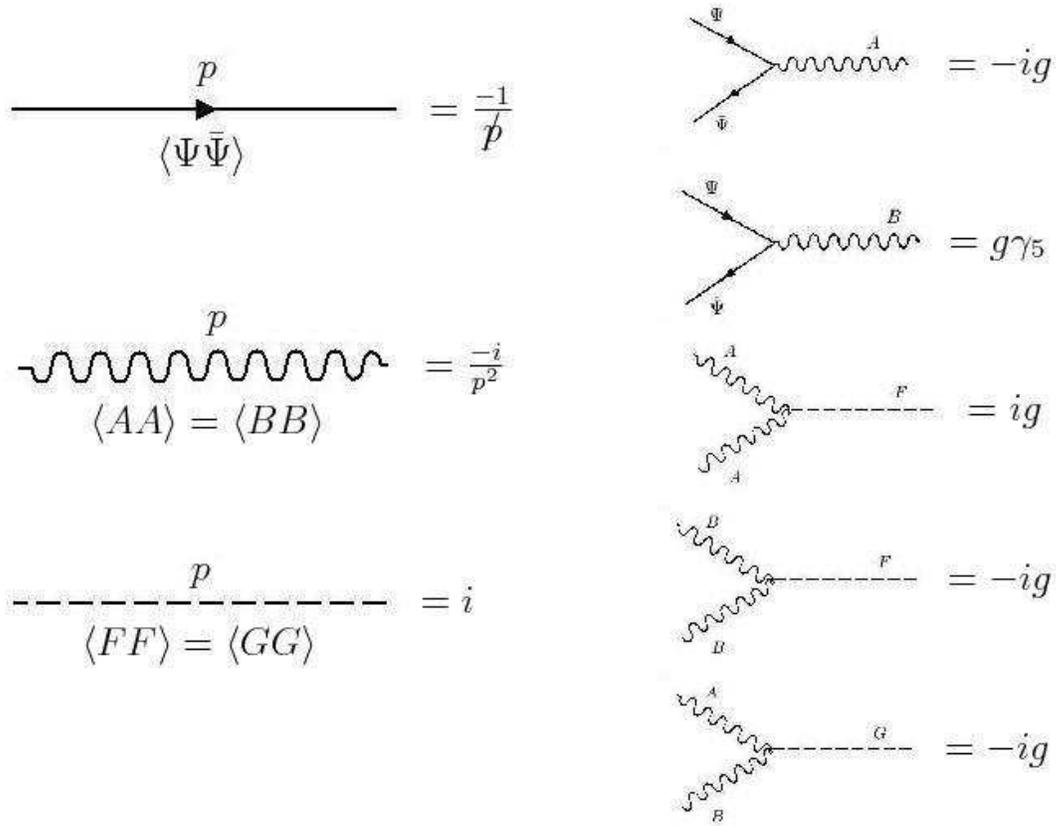}
\end{center}

\caption{\label{figfrwz} Feynman rules for the massless Wess-Zumino model.}
\end{figure}


\begin{thebibliography}{99}

\bibitem{thooftveltman} G. 't Hooft and M. J. G. Veltman, \textit{Nucl. Phys.
\textbf{B}} \textbf{44} (1972) 189.

\bibitem{MCKEON} V. Elias, G. McKeon, S. B. Phillips, R. B. Mann, \textit{Phys. Lett. B} \textbf{133} (1983) 83.

\bibitem{Stockinger} D. Stockinger,\textit{Nucl. Phys. Suppl.} \textbf{160} (2006)
250; \textit{JHEP} \textbf{0503} (2005) 076.

\bibitem{Jackiw:1999qq}
  R.~Jackiw,
  Int.\ J.\ Mod.\ Phys.\  {\bf B14}, 2011-2022 (2000).

\bibitem{JACKIWCURRENTALGEBRA} R. Jackiw, \textit{hep-th/0011274}.

\bibitem{Mota} O. A. Battistel, A. L. Mota, M. C. Nemes \textit{{Mod.
Phys. Lett. \textbf{A}}} \textbf{13} (1998) 1597.

\bibitem{Bogoliubov} N. N. Bogoliubov and O. S. Parasiuk, {\it{Introduction to the theory of quantized
fields}}, John Wiley (1980); E. B. Manoukian,
{\it{Renormalization}}, Academic Press (1983).

\bibitem{Carlos2} C. R. Pontes, A. P. Ba\^eta Scarpelli, Marcos Sampaio and M. C. Nemes
\textit{J. Phys. G} \textbf{34}, (2007) 2215.

\bibitem{Edson} E. W. Dias, A. P. Ba\^eta Scarpelli, L. C. T. Brito, Marcos Sampaio, M. C. Nemes,
\textit{Eur. Phys. J. } \textbf{C} \textbf{55} (2008) 667.

\bibitem{Cleber} L. C. T. Brito, H. Fargnoli, A. P. Ba\^eta
Scarpelli, M. Sampaio, M. C. Nemes \textit{Phys. Lett. B}
\textbf{673}, (2009) 220.

\bibitem{Adriano}
  A.~L.~Cherchiglia, M.~Sampaio, M.~C.~Nemes,
\textit{ Int. J. Mod. Phys.} {\bf A26 } (2011)  2591-2635.

\bibitem{BJP} E. W. Dias, A. P. Ba\^eta Scarpelli, L. C.
Brito, H. G. Fargnoli, \textit{Braz. J. Phys.} \textbf{40, n 2} (2010) 228.

\bibitem{BETAYM} H. G. Fargnoli, B. Hiller, A. P. Ba\^eta Scarpelli, M. Sampaio, M. C. Nemes,
\textit{Eur.Phys. J. C.} \textbf{71} (2011) 1633.

\bibitem{Prd1}  A. P. Baeta Scarpelli, M. Sampaio and M. C. Nemes,
\textit{{Phys. Rev. \textbf{D}}} \textbf{63} (2001) 046004.

\bibitem{Prd2}  A. P. B. Scarpelli, M. Sampaio, M.
C. Nemes and B. Hiller, \textit{{Phys. Rev. \textbf{D}}} \textbf{64} (2001) 046013.

\bibitem{David} D. Carneiro, A. P. Ba\^eta Scarpelli, M. Sampaio and M. C.
Nemes,

\textit{JHEP} \textbf{12} (2003) 044.

\bibitem{Eloy2} J. E. Ottoni, A. P. Ba\^eta Scarpelli, Marcos Sampaio, M. C. Nemes, \textit{Phys. Lett.} \textbf{B 642} (2006) 253.

\bibitem{DiffRen2} F. del Aguila, A. Culatti, R. Mu\~noz Tapia and M.
P\'erez Victoria, \textit{{Phys. Lett. \textbf{B}}} \textbf{419},
263 (1998);\textit{{Nucl. Phys. \textbf{B}}} \textbf{504}, 532
(1997);\textit{{Nucl. Phys. \textbf{B}}} \textbf{537}, 561 (1999).


\bibitem{Carlos1} C. R. Pontes, A. P. Ba\^eta Scarpelli, Marcos Sampaio, J. L.
Acebal, M. C. Nemes, \textit{Eur. Phys. J. C}\textbf{53}(2008)121.

\bibitem{Seijas} C. Seijas, \textit{Ann. Phys.} \textbf{322} (2007) 1972.

\bibitem{Abbott} L. F. Abbott, M. T. Grisaru, D. Zanon, \textit{Nuc. Phys.} \textbf{B 244} (1984) 454.

\bibitem{Delamotte} B. Delamotte, \textit{Am. J. Phys.} \textbf{72} (2004) 170.

\bibitem{PerezVictoria:2001ej}
  M. Perez-Victoria,
  {\it J. High Energy Phys.} {\bf 0104} (2001) 032.

\bibitem{Peskin}
  Michael E. Peskin, Dan V. Schroeder
  {\it ``An Introduction To Quantum Field Theory"},
	Addison-Wesley Publishing Company (1995).
	
\bibitem{Bertlmann}
  R.~A.~Bertlmann,
  Oxford, UK: Clarendon (1996) 566 p. (International series of monographs on physics: 91)	   	

\bibitem{wess} J. Wess and B. Zumino, Phys. Lett. B 49, 52 (1974).

\bibitem{Eloy}
  M.~D.~Sampaio, A.~P.~Baeta Scarpelli, J.~E.~Ottoni and M.~C.~Nemes,
  Int.\ J.\ Theor.\ Phys.\  {\bf 45}, 436 (2006)
  
\bibitem{quantumaction}
  A. L. Cherchiglia, M. C. Nemes, B. Hiller, M. Sampaio,
  work in progress  
  

\end{thebibliography}
\end{document}